# Hybrid graphene-quantum dot phototransistors with ultrahigh gain

Gerasimos Konstantatos[1,*], Michela Badioli[1], Louis Gaudreau[1], Johann Osmond[1], Maria Bernechea[1], F. Pelayo Garcia de Arquer[1], Fabio Gatti[1], Frank H. L. Koppens[1,*]

[1] ICFO - Institut de Ciències Fotòniques, Mediterranean Technology Park, 08860 Castelldefels, Barcelona, Spain

Corresponding authors: gerasimos.konstantatos@icfo.es, frank.koppens@icfo.es

**Graphene has emerged as a novel platform for opto-electronic applications and photodetection [1-14], but the inefficient conversion from light to current has so far been an important roadblock. The main challenge has been to increase the light absorption efficiency and to provide a gain mechanism where multiple charge carriers are created from one incident photon. Here, we take advantage of the strong light absorption in quantum dots and the two-dimensionality and high mobility of graphene to merge these materials into a hybrid system for photodetection with extremely high sensitivity. Exploiting charge transfer between the two materials, we realize for the first time, graphene-based phototransistors that show ultrahigh gain of $10^8$ and ten orders of magnitude larger responsivity compared to pristine graphene photodetectors. These hybrid graphene-quantum dot phototransistors exhibit gate-tunable sensitivity, spectral selectivity from the shortwave infrared to the visible, and can be integrated with current circuit technologies.**

Graphene, a single atomic-layer of carbon, has recently been introduced as a novel two-dimensional material presenting extraordinary electronic, structural, mechanical and optical properties [1-5]. In particular, graphene photodetectors have received considerable attention because of the broad spectral bandwidth and ultrafast response time [6-10]. However, the responsivity (the photogenerated current per incident optical power ) is less than $10^{-2}$ A/W [6-10], limited by the weak light absorption in graphene. While the absorption of light in graphene can be improved by exploiting thermo-electric effects [11-13], metallic plasmonics [14] or graphene plasmons [15], the key requirement for ultra-sensitive graphene-based photodetection is the implementation photoconductive gain - the ability to provide multiple electrical carriers per single incident photon. However until now gain has not been observed in graphene-based photodetectors.

Photodetector gain for ultrahigh sensitivity has been exploited in avalanche photodiodes and photomultipliers; however these devices require application of high electric bias, are bulky and hence are not compatible with current circuit technologies, e.g. CMOS. Phototransistors have also been reported to provide gain at low temperature for visible wavelengths [16] or for the short-wave infrared (SWIR) [17], yet this has been achieved on III-V semiconductors with limited potential for CMOS integration. Recently, photoconductive gain was observed in colloidal quantum dot (QD) photodetectors, which have emerged as top-surface detectors offering integration with circuit technologies in view of their solution-processability. These low-cost high sensitivity detectors operate in the visible and short-wave infrared, tunable by the bandgap of the quantum dots [18] These colloidal QD photodectors have shown gain of $\sim 10^2$-$10^3$ charge carriers per incident photon [19, 20], predominantly limited by the relatively poor carrier mobility of colloidal quantum dots ($10^{-3} - 1$ $cm^2V^{-1}s^{-1}$).

The charge carriers in graphene, on the contrary, exhibit very high mobility (up to 60.000 $cm^2$/Vs on a substrate and at room temperature [21]) and since graphene is an intrinsically two-dimensional material, its conductance is very sensitive to electrostatic perturbation by photogenerated carriers close to the surface. As we show below, this makes graphene a particularly promising material for high gain photodetection by employing the photogating effect. Additionaly, graphene is a very thin, flexible and durable material which can be fabricated in a large scale [22-25] and easily deposited on silicon, offering monolithical integration to standard integrated circuits. Thus, the demonstration of photodetection gain with graphene would be the basis for a plethora of applications such as graphene-based integrated opto-electronic circuits, biomedical imaging, remote sensing, optical communications, and quantum information technology.

Here, we present a novel hybrid graphene-quantum dot phototransistor which exhibits ultrahigh photodetection gain and high quantum efficiency, enabling high-sensitivity and gate-tunable photodetection. The key functionality of this light-activated transistor is provided by a layer of strongly light-absorbing and spectrally tunable colloidal quantum dots, from which photo-generated charges can transfer to graphene, while oppositely charged carriers remain trapped in the QD layer. These trapped carriers lead to a photogating effect, where the presence of these charges changes the graphene sheet resistance trough capacitative coupling (Fig. 1A). Our hybrid phototransistor devices reveal quantum efficiencies exceeding 25%, and photoconductive gain of $10^8$ carriers per absorbed photon for a bias of 1V, corresponding to a responsivity of ~$10^8$ A/W, which is ten orders of magnitude larger than the responsivity of graphene photodetectors reported so far [6-10,13].

Graphene's unique electronic properties offer gate-tunable carrier density and polarity which enables us to tune the sensitivity and operating speed of the detector. We expoit this to maximize the photoconductive gain or to fully reduce it to zero, which is useful for pixelated imaging applications, where the implementation of nanoscale local gates enables a locally tunable photoresponse. Moreover, we show that a short voltage pulse applied to the gate can be used to actively purge the charge carriers from the quantum dots in order to reset the device and thereby increase the operating speed of the device.

The device consists of a graphene sheet that is sensitized with colloidal quantum dots. Graphene is the carrier transport channel, and the quantum dots are employed as the photon absorbing material (Fig 1A). The channel of the phototransistor, consisting of a monolayer or bilayer graphene sheet, is placed atop a Si/SiO$_2$ wafer, as shown in the inset of Figure 1B (see supplementary information for details of device fabrication procedures). Gold contacts define the source and drain, and the graphene flake can be electrically gated using the backgate of the Silicon wafer (see SOM for electrical characterization before and after QD deposition). The channel fabrication was followed by a deposition of thin films of PbS colloidal quantum dots from solution via spincasting. We employed the layer-by-layer (LBL) approach to build a ~80 nm thick film of PbS quantum dots with the first exciton peak at ~950 nm or ~1450 nm. We thus demonstrate two types of phototransistors with sensitivities up to 1050 nm and 1600 nm. During the LBL the quantum dots underwent a ligand exchange process to replace the long oleate ligands from their surface to shorter bidentate ligands of ethanedithiol in order to turn the quantum dots into a conductive solid state film [26]. Figure 1B shows the spatially resolved photocurrent response of our structure upon illumination, measured with a focused laser beam (wavelength=532 nm, spotsize ~500 nm). The photocurrent was recorded as the laser beam was scanned across the surface of the detector. Contrary to what has been shown previously in graphene-based detectors, where photocurrent generation occurs at the interface of graphene with metal contacts [6-10], or in the vicinity of a pn-junction [11,12,27], our structure is photo-responsive over a large area, a feature of importance in most sensing applications.

In Figure 1C we plot the spectral responsivities of two devices that contain PbS QDs which have their first exciton peaks either at ~950 nm (top panel) or at ~1450 nm (bottom panel). Clearly, the photocurrent response follows the absorption of the PbS QDs, whereas no photocurrent is observed for photon energies below the bandgap of the QD layer. Photocarrier generation at the graphene layer is not expected to yield photoconductance due to the ultrafast recombination in graphene [28,29]. These measurements therefore demonstrate spectral selectivity of our graphene-quantum dot phototransisters, offered by the bandgap tunability of the quantum dots. Remarkably, at low excitation power, we measure a responsivity as high as $~10^8$ $AW^{-1}$, corresponding to a photoconductive gain of $0.5*10^8$ for an excitation wavelength of 600 nm.

We turn now to the specifics of the device and describe the physical mechanism underlying the observed photocurrent response. We first measured the source-drain resistance as a function of backgate voltage under dark and ambient equilibrium conditions. The dramatic shift of the Dirac point from ~120 V to ~50 V upon deposition of the quantum dot layer (see Fig. S3) shows that holes from graphene are transferred to the quantum dot layer, forming a built-in field at the in order to equilibrate the Fermi levels (as shown in Fig. 2D). We then monitored the resistance of graphene under illumination (in vacuum conditions) using a collimated laser beam of about 1 mm diameter, while changing the Fermi energy with the backgate. Figure 2A shows the resistance R as a function of backgate voltage $V_{BG}$ under different illumination levels, revealing the typical Dirac peak with the Dirac point $V_D$ where the resistance is maximum. We observe that illumination causes the Dirac point to shift to higher values of $V_{BG}$, and thus the resistance of the graphene channel decreases for $V_{BG} < V_D$, where the carrier transport is hole dominated whereas it increases for $V_{BG} > V_D$, where carrier transport is electron dominated. This photo-induced shift of $V_D$ is present over a four orders of magnitude range of optical intensities (Fig. 2A, inset).

From these data, we extract the shift of $V_D$ as a function of optical intensity (Fig. 2B) which clearly reveals the photosensitivity of our device: for low laser power the light-induced shift of $V_D$ can be as large as 20 V/pW (Fig. 2B) enabling the detection of laser powers smaller than 10 fW. The dependence of the photosignal on bias is shown in Fig 2C, which reveals that the photocurrent ($\Delta I = I_{Light} - I_{Dark}$) depends linearly on the bias voltage $V_{SD}$.

These observations lead to the following model of the underlying physical mechanism of photocurrent in the device: photon absorption in PbS QDs creates electron hole pairs in the QD film which then separate at the interface of graphene with the QDs. This charge separation is induced by an internal electric field that leads to band bending at the interface due to the work function mismatch between graphene and the QDs, illustrated in Fig 2D. For $V_{BG}<V_D$, graphene is hole doped, and illumination causes a decrease in the resistance. This points to photogenerated holes that are transferred from the QDs to graphene. For $V_{BG}>V_D$, illumination leads to resistance increase or current quenching, which is due to recombination that takes place between photogenerated holes transferred from the QDs to graphene and electrons induced by the backgate. As the photogenerated electrons remain trapped in the QD phase, the negatively charged quantum dots induce positive carriers in the graphene sheet through capacitive coupling, which explains the observed shift of the Dirac peak towards higher backgate voltages. As long as the QDs remain negatively charged, positive charges in the graphene sheet are recirculated, resulting in gain. This photogating effect is schematically shown in Fig. 2D.

We note that the photo-induced shift of the Dirac peak can be reversed, depending on the initial doping level of the graphene sheet. Thus, an initially negatively doped graphene sheet may form a built-in potential at the interface with the PbS QDs such that photogenerated electrons are transferred to the graphene sheet and the holes remain trapped in the PbS QD layer (see SOM). This is of importance as the initial doping of the graphene can thus be used as an additional leverage to develop phototransistors employing a variety of semiconductor QDs with different spectral sensitivities.

Since the device behavior, and in particular the shift of $V_D$, can be understood by considering the negatively charged quantum dot layer as a local photoinduced gate that modifies the graphene carrier concentration through capacitive coupling, we can use a simple parallel plate capacitor model to estimate the internal quantum efficiency QE of the device. This model is valid for a QD thickness that is much smaller than the size of the graphene flake. Under this condition, the charge carrier density in the graphene should be equal to the charge carrier density in the quantum dots, but with opposite polarities. We find that 0.1 pW laser power (corresponding to a photon flux of $\phi_{photon} =$

$6*10^{11}$ cm$^{-2}$s$^{-1}$) induces a 2 V shift of $V_D$, implying that during one second (see discussion below on $\tau_{lifetime}$) a negative charge carrier density of about $1.5*10^{11}$ cm$^{-2}$ is induced (for 285 nm thick oxide, the capacitive coupling of the backgate is $7*10^{10}$ cm$^{-2}$V$^{-1}$). This implies that about 25% of the incident photons are converted to negatively trapped carriers in the quantum dots that contribute to the photogating effect. Thus, our devices, yet unoptimized, show an external quantum efficiency of 25%.

In Fig. 3A we plot the responsivity of our device as a function of the applied backgate voltage. Taking into account the active device area, we find that the responsivity is orders of magnitude greater than unity. We measure a responsivity as high as $10^7$ AW$^{-1}$ when $V_{BG} < V_D$, corresponding to a photoconductive gain of $0.5*10^8$ for an excitation wavelength of 600 nm. This gain is an improvement of 9-10 orders of magnitude compared to pristine graphene and 5-6 orders of magnitude compared to earlier QD-based photodetectors [19, 20]. By tuning the Fermi energy close to the Dirac point at $V_{BG} = 4$ V, the responsivity completely falls to 0. This feature demonstrates the potential of this device as a backgate-tunable ultrahigh-gain phototransistor. This tunability is of high importance in photodetectors because it allows the control of the state (ON-OFF) of the detector as well as the adjustment of the required gain, depending on the light intensity to be detected.

We also exploit the back-gate tunability of the graphene Fermi level to develop a reset functionality in our detectors as an electronic shutter suited to video-frame-rate imaging applications. In Fig. 3B, we plot the temporal response of the photodetector at different power levels. The time response of the photocurrent decay is dominated by two components with faster lifetimes of about 10-20 ms (corresponding to 50% decay) and a slow component of 2 s (Fig. 3B, inset), likely associated with the multiplicity of electron traps in PbS QDs from different surface states [30]. The bottom panel of Fig. 3B shows the accelerated photocurrent decay upon an electric pulse applied at the gate of the device. This pulse generates an electric field that reduces the potential barrier which keeps electrons trapped in the quantum dots at the graphene-QD interface. As the potential barrier is reduced, trapped electrons can escape to the graphene sheet followed by a rapid decay of photocurrent (Fig. 3B, bottom). The temporal response of the detector can thus be accelerated to ~20 ms allowing for e.g. video rate imaging applications (see section 5 of SOM).

The main feature of the devices is the ultrahigh gain, which originates from the high carrier mobility of the graphene sheet ($\sim 10^3$ $cm^2V^{-1}s^{-1}$) and the recirculation of charge carriers through the prolonged lifetime of the carriers that remain trapped in the PbS QDs [30]. Photoexcited holes in the PbS QDs are transferred to the graphene layer and drifted by $V_{DS}$ to the drain, with a typical timescale of $\tau_{transit}$, which is inversely proportional to the carrier mobility. Electrons remain trapped with a typical timescale of $\tau_{lifetime}$ in the PbS QDs due to the built-in field at the QD-graphene interface (Fig. 2D) as well as the electron traps in PbS QDs. Charge conservation in the graphene channel leads to hole replenishment from the source as soon as a hole reaches the drain. Therefore multiple holes circulate the graphene channel, upon a single electron-hole photogeneration, leading to photoconductive gain. The photoconductive gain for this mechanism is given by $G = \tau_{lifetime} / \tau_{transit}$ [31], evidencing the importance of a long lifetime and a high carrier mobility. To verify that our experimentally observed gain is in agreement with the theoretical gain, we determine $\tau_{lifetime}$ and $\tau_{transit}$. For the specific device, the transit time of the carriers is found to be on the order of 1 ns (based on the extracted mobility $\mu$ of $10^3$ cm$^2$V$^{-1}$s$^{-1}$ from the data of Fig. 2A, a channel length of 10 μm and applied bias $V_{DS}$ of 1 V), which then leads to a predicted range of values for the photoconductive gain between $10^7$ and $10^9$, using lifetimes of 20 ms and 1 s respectively, in good agreement with measured gain values. The device based on bilayer graphene exhibits a smaller responsivity, which can be explained by its smaller mobility of about 600 cm$^2$V$^{-1}$s$^{-1}$ and the lower built-in field at the interface with the PbS QDs (see SOM device 2). We remark that a simple alternative relation for the photoconductive gain can be obtained based on the change in conductivity $\Delta\sigma=\Delta n*e*\mu$ due to the light-induced modification of the graphene carrier concentration $\Delta n= \tau_{lifetime}*QE*\phi_{photon}$. This gives a theoretical prediction for the gain $G = \tau_{lifetime}*QE*\mu*V_{SD}/A$, with $A$ the area of the graphene flake. For our devices, we find $G=2*10^8$, in excellent agreement with the experimental observations and the (physically equivalent) model discussed above. The obtained relation for the gain gives the insight that the gain can be further improved by at least two orders of magnitude by increasing the bias and improving the graphene carrier mobility.

In order to measure the sensitivity of the detector we performed differential resistance and responsivity measurements under variable optical intensity. Fig. 3C (upper panel) shows the responsivity as a function of the incident optical power on the detector area. The responsivity remains ~$10^8$ AW$^{-1}$ for optical power up to 50 fW followed by a decrease with increasing light intensity. This photocurrent saturation effect is due to the lowering of the built-in field at the QD-graphene interface with increasing number of photogenerated electrons which generate an electric field opposed to the equilibrium built-in field. This phenomenon may be of particular importance in high dynamic range sensing for avoiding saturation effects at the read-out at intense illumination levels.

The lower panel of Fig. 3C shows a direct measurement of 8 fW of incident light which demonstrates the ultra-high sensitivity of the device. In order to determine the ultimate sensitivity of the phototransistor we measured the noise level floor of the transistor under dark conditions (of $0.1\Omega.Hz^{-1/2}$, see supporting information for more details) which sets the lower threshold limit of noise equivalent power of $10^{-17}$ W (10 aW). This yields a specific detectivity D* of $7 \cdot 10^{13}$ Jones ($cmHz^{1/2}W^{-1}$) which is the highest reported among any quantum dot or graphene photodetector to date. This value is also an order of magnitude greater than D* achieved in Si and InGaAs detectors currently used in high sensitivity imaging applications.

Significant improvements can be realized by the implementation of nanopatterned graphene or bilayer graphene where the opening of a band-gap would lead to enhanced sensitivity and suppression of dark current. By combining our sensitization technique with graphene-based single electron detectors [32, 33], the integration of opto-electronic circuits with graphene-based photodetectors sensitive to very small photon numbers is within reach. On a broad perspective we have developed the first successful hybrid material platform of graphene with another material to demonstrate a photodetector with performance superior to what has been achieved before by the individual technologies. Our study shows that efficient electronic coupling of graphene with other technologies such as light-absorbing materials will open pathways for novel optoelectronic functionalities and energy harvesting applications.

**Figure Captions:**

**Fig. 1. (A)** A schematic of the graphene-QD hybrid phototransistor: a graphene flake is deposited onto a $SiO_2$/Si structure overcoated with PbS QDs. Incident photons create electron-hole pairs in PbS QDs. Holes are then transferred to the graphene channel and drift towards the drain whereas electrons remain in the PbS QDs, where through capacitive coupling they lead to a prolonged time during which (recirculated) carriers are present in the graphene channel. **(B)** Spatial photocurrent profile using a focused laser beam at 532 nm and power=1.7 pW. The spatial profile shows the large area excitation of the phototransistor at the overlapping area of the quantum dot film with the graphene flake. $V_{SD}$=10 mV. Inset: optical image of the graphene flake used in this study in contact with the Au electrodes to form the phototransistor. **(C)** Spectral responsivity of two phototransistors of respectively single- and bilayer graphene, employing PbS QDs of different sizes with exciton peaks at either 950 nm (top panel) or 1450 nm (bottom panel). The spectral sensitivities of the devices are determined by the absorption spectra of the QDs and can be easily tuned via control of the QD size (lines are drawn as guides to the eye).

**Fig. 2. (A)** Resistance as a function of backgate voltage for the graphene-QD structure for increasing illumination intensities using a collimated laser with diameter of 1 mm and wavelength of 500 nm. Increasing illumination leads to a photogating effect that shifts the Dirac point to higher backgate voltage. This indicates hole photo-doping of the graphene flake. Inset: Two-dimensional plot of the the graphene resistance as a function of optical power (same parameters as in (A)) **(B)** Shift of the Dirac point as a function of optical power that illuminates the graphene flake. **(C)** Photocurrent of the graphene-QD transistor for different optical powers as a function of the drain-source voltage ($V_{DS}$) shows a linear dependence on bias ($V_{BG}$=0). The inset shows the total current ($I_{dark}+I_{photo}$). **(D)** Energy level diagram of the graphene-QD interface: Upon deposition of QDs on graphene, the Dirac point of graphene shifts from 120 V to 50 V, indicating a hole transfer from the graphene to the QDs (see SOM). A built-in field is thus formed at the interface as shown in the top-panel, illustrating the bands for $V_{BG}=V_D$. Upon photoexcitation of PbS QDs holes are transferred to the graphene under the built-in field, leaving electrons trapped in PbS QDs. This lowers the Fermi energy in the graphene (bottom panel).

**Fig. 3.** Phototransistor device characteristics. **(A)** Responsivity as function of the backgate shows that the gain of the detector can be directly controlled by the applied backgate potential and can be tuned from $4.10^7$ ($V_{BG}$ = -20 V) to 0 ($V_{BG}$ = 4 V). $V_{SD}$=5 V. **(B)** Top panel: Temporal photocurrent response of the device for wavelength 532 nm and power= pW . The temporal response indicates a rise-time of about 10ms and two different fall-times on the order 100 ms (50%) and 1 s (see inset, measured at higher power=267 pW). Bottom panel: temporal response of a bilayer graphene phototransistor

after the laser is turned off and application of a reset pulse during 10 ms. The fall-time is reduced from several seconds to ~10ms (see SOM) **(C)** Responsivity and light-induced resistance change (ΔRes) versus optical illumination power shows the high sensitivity of the phototransistor of 8 fW by direct measurement. The transistor exhibits responsivity of $6.10^7$ $AW^{-1}$ for optical power less than 50 fW and shows progressive decrease with increasing illumination as consequence of the reduction of the built-in field with increasing number of photogenerated carriers. The solid line is the best fit to the data using the function $R=c_1/(c_2+P)$, with $c_1$ and $c_2$ the fit parameters. The lower threshold limit of noise equivalent power of $10^{-17}$ W is set by the noise floor of $0.1\Omega.Hz^{-1/2}$.

**References**

(1) K.S. Novoselov *et al, Science* **306,** 666 (2004)

(2) K. S. Novoselov *et al, Nature* **438**, 197 (2005)

(3) Y. Zhang, Y.W. Tan, H.L. Stormer, Ph. Kim. *Nature* 438, 201-204 (2005)

(4) A. Geim, *Science* **324,** 1530-1534 (2009)

(5) F. Bonaccorso, Z. Sun, T. Hasan, A. C. Ferrari, *Nature Phot.* **4**, 611 (2010).

(6) J. Park, Y. H. Ahn, C. Ruiz-Vargas, *Nano Lett.* **9**, 1742 (2009).

(7) E. J. H. Lee, K. Balasubramanian, R. T. Weitz, M. Burghard, K. Kern, *Nature Nano.* **3**, 486 (2008)

(8) F. Xia, *et al. Nano Lett.* **9**, 1039 (2009).

(8) *F. Xia., T. Mueller., Y. M. Lin, A. Valdes-Garcia P. Avouris,* Nature Nanotechnology 4, 839 - 843 (2009).

(10) T. Mueller, F. Xia, P. Avouris, *Nature Phot.* **4**, 297-301 (2010).

(11) M. Lemme et al, *Nano Lett.,* 11 (10), pp 4134–4137 (2011).

(12) N. M. Gabor et al, *Science* DOI: *10.1126/science.1211384* (2011)

(13) Justin C. W. Song, Mark S. Rudner, Charles M. Marcus, and Leonid S. Levitov, *Nano Lett.,* **DOI:** 10.1021/nl202318u (2011).

(13) T.J. Echtermeyer et al. *Nature Comm.,* 2, Article number: 458, doi:10.1038/ncomms1464 (2011).

(14) F.H.L. Koppens, D. E. Chang, F. J. García de Abajo Nano Lett. **11**, 3370–3377 (2011)

(15) S. Thongrattanasiri, F. H. L. Koppens, F. Javier Garcia de Abajo, Arxiv 1106.4460.

(16) A. J. Shields et al, *Appl. Phys. Lett*. 76, 3673 (2000).

(17) P. D. Wright, R. J. Nelson, T. Cella, *Appl. Phys. Lett.*, 37, 192 (1980)

(18) G. Konstantatos, E. H. Sargent, *Nature Nanotechnol.* **5**, 391–400 (2010)

(19) G. Konstantatos et al, *Nature,* 442, 180 (2006)

(20) J. S. Lee, M. V. Kovalenko, J. Huang, D. S. Chung, D. V. Talapin, *Nature Nano.* 6, 348 (2011)

(21) C. R. Dean, A. F. Young, I. Meric, C. Lee, L. Wang, S. Sorgenfrei, K. Watanabe, T. Taniguchi, P. Kim, K. L. Shepard and J. Hone, *Nature Nano.* 5, 722-726 (2010)

(22) Y. Lee *et al. Nano Letters* **10**, 490(2010)

(23) A. Reina *et al. Nano Letters* **9**, 30 (2009)

(24) X. Li et al, *Science,* Vol. 324, 5932, 1312 (2009)

(25) S. Bae et al. Nature Nanotechnology 5, 574–578 (2010)
(26) J. M. Luther et al, *ACS NANO,* 2, 271 (2008)

(27) E. C. Peters et al., *Appl. Phys. Lett.* **97**, 193102 (2010)

(28) P.A. George et al., Nano Lett 8, 4248–4251 (2008).

(29) Dawlaty, J., Shivaraman, S. & Strait, *Appl. Phys. Lett.* **93,** 131905 (2008)

(30) G. Konstantatos, L. Levina, A. Fischer, E. H. Sargent, *Nano Lett.* 8, 1446-1450 (2008)

(31) A. Rose, Concepts in photoconductivity and allied problems, Robert E. Krieger Publishing Co. (1978)

(32) F. Schedin et al., *Nature Mater.* **6**, 652–655 (2007)

(33) J. Moser, A. Bachtold, *Appl. Phys. Lett.* **95**, 173506 (2009)

Acknowledgements:

This research has been partially supported by Fundació Cellex Barcelona. We thank Klaas-Jan Tielrooij for fruitful discussions.

Figure 1

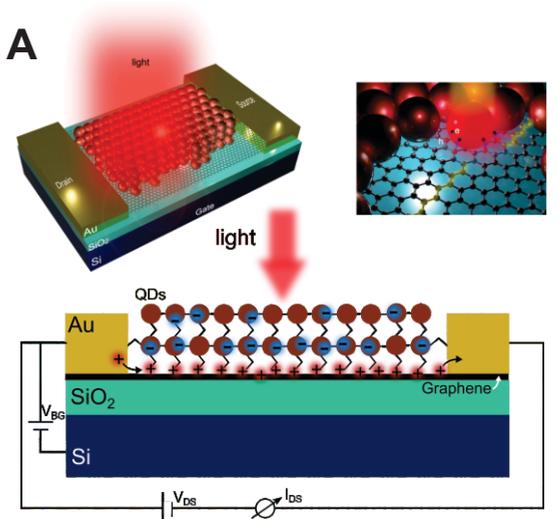

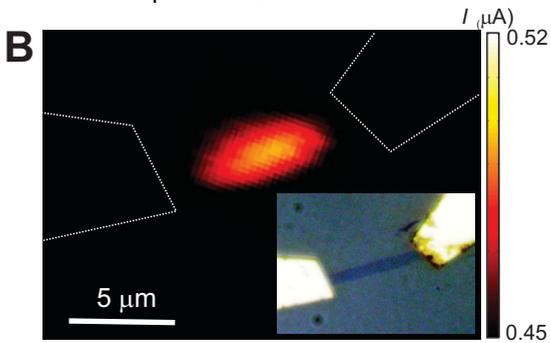

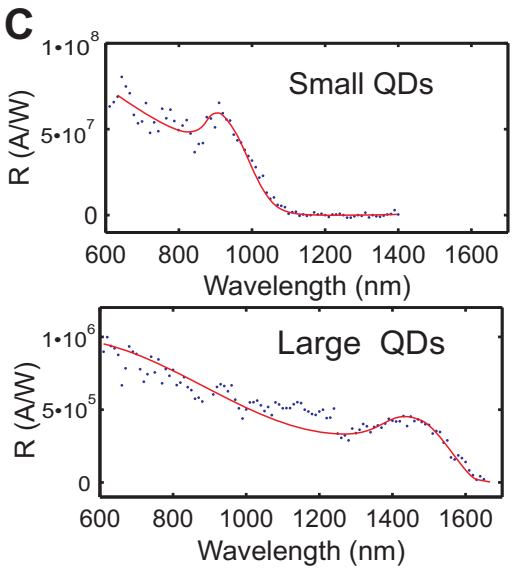

# Figure 2

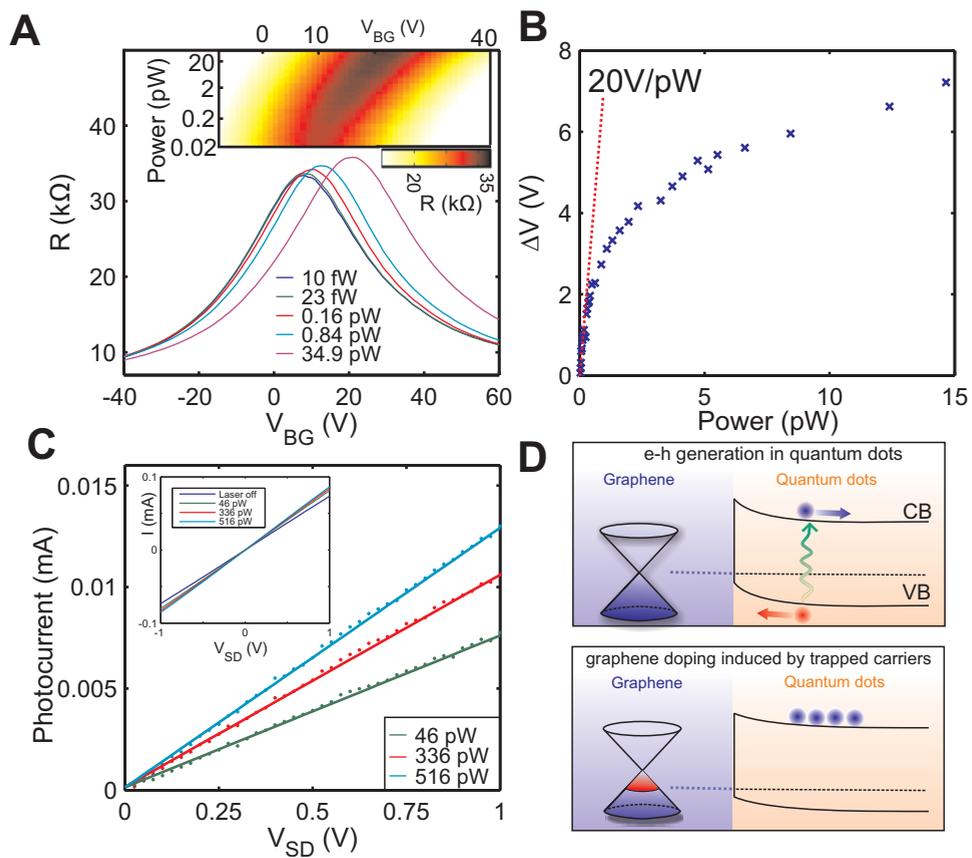

# Figure 3

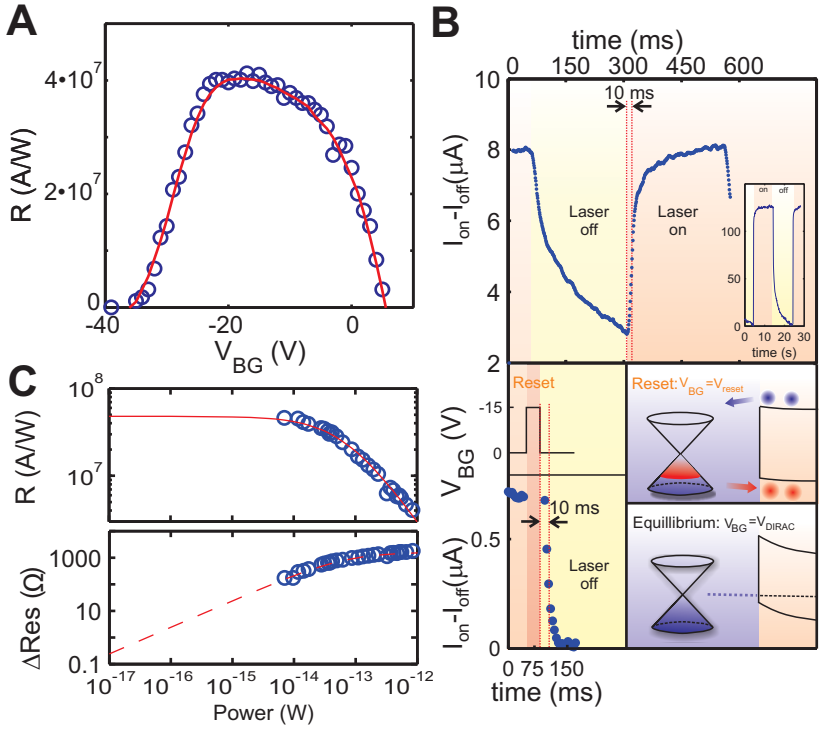